\documentclass[twocolumn]{aastex631}
\usepackage{color}

\newcommand{\lya}{\hbox{Ly$\alpha$}}

\newcommand{\hb}{\hbox{H$\beta$}}

\newcommand{\gsim}{\lower.5ex\hbox{$\; \buildrel > \over \sim \;$}}
\newcommand{\lsim}{\lower.5ex\hbox{$\; \buildrel < \over \sim \;$}}

\newcommand{\oii}{\hbox{[O\,{\sc ii}]}}
\newcommand{\oiii}{\hbox{[O\,{\sc iii}]}}

\newcommand{\civ}{\hbox{C\,{\sc iv}}}
\newcommand{\heii}{\hbox{He\,{\sc ii}}}
\newcommand{\ciii}{\hbox{C\,{\sc iii}]}}
\newcommand{\oiiij}{\hbox{O\,{\sc iii}]}}
\newcommand{\feii}{\hbox{Fe\,{\sc ii}}}
\newcommand{\siii}{\hbox{Si\,{\sc ii*}}}

\newcommand{\siiii}{\hbox{Si\,{\sc ii}}}

\begin{document}

\title{Galactic-scale emission-line outflow from the radio-loud quasar 3C 191}

\author{Qinyuan Zhao}
\affiliation{Department of Astronomy, University of Science and Technology of China, Hefei 230026, China} 
\email{Corresponding authors: Qinyuan Zhao (zqy94070@mail.ustc.edu.cn), Guilin Liu (glliu@ustc.edu.cn)}

\affiliation{School of Astronomy and Space Sciences, University of Science and Technology of China, Hefei 230026, China}
\affiliation{Department of Astronomy, Xiamen University, Xiamen, Fujian 361005, China}

\author{Luming Sun}
\affiliation{Department of Physics, Anhui Normal University, Wuhu, Anhui 241002, China}

\author[0000-0001-9495-7759]{Lu Shen}
\affiliation{Department of Physics and Astronomy, Texas A\&M University, College Station, TX, 77843-4242 USA}
\affiliation{George P.\ and Cynthia Woods Mitchell Institute for Fundamental Physics and Astronomy, Texas A\&M University, College Station, TX, 77843-4242 USA}

\author[0000-0003-4286-5187]{Guilin Liu}
\affiliation{Department of Astronomy, University of Science and Technology of China, Hefei 230026, China}
\affiliation{School of Astronomy and Space Sciences, University of Science and Technology of China, Hefei 230026, China}

\author[0000-0003-4874-0369]{Junfeng Wang}
\affiliation{Department of Astronomy, Xiamen University, Xiamen, Fujian 361005, China}
\author[0009-0001-5990-5790]{Mayank Sharma}
\affiliation{Department of Physics, Virginia Tech, Blacksburg, VA 24061, USA}
\author[0000-0003-2991-4618]{Nahum Arav}
\affiliation{Department of Physics, Virginia Tech, Blacksburg, VA 24061, USA}
\author{Yulong GAO}
\affiliation{School of Astronomy and Space Science, Nanjing University, Nanjing 210093, PR China}
\affiliation{Key Laboratory of Modern Astronomy and Astrophysics (Nanjing University), Ministry of Education, Nanjing 210093, PR China}
\author{Chris Benn}
\affiliation{}



\begin{abstract}

Quasar feedback is routinely invoked as an indispensable ingredient in galaxy formation models. Galactic outflows are a crucial agent of quasar feedback that frequently manifest themselves in absorption and emission lines. Measuring the size and energetics of outflows based on absorption lines remains a challenge, and integral-field spectroscopy (IFS) mapping in emission lines is complementary. We present a VLT/SINFONI IFS mapping of quasar 3C 191 at $z \sim 2$, in which the outflow has been analyzed in absorption line spectroscopy. Three components are found based on the morphology and kinetics of \oiii-emitting gas: a unshifted component which consistent with the systemic redshift and the location of the nucleus, a blueshifted in the north, and a redshifted in the south. The latter two components have velocities $\sim$ 600 km s$^{-1}$ and projected extents of 5 and 11 kpc, respectively, suggesting a biconical outflow structure. The blueshifted component’s velocity is 
consistent with that derived from absorption lines. Using the electron density measured by the absorption lines and the luminosity and velocity of \oiii\ outflow, we derive the mass outflow rate to be $\dot{M} \sim $ 9.5-13.4 M$_\odot$ yr$^{-1}$ and kinetic luminosity $\dot{E}_{\rm kin} \sim$ 2.6-3.7 $\times 10^{42}$ erg s$^{-1}$, consistent with absorption line analyses with VLT/Xshooter spectrum. The kinetic luminosity is only 0.01\% of the bolometric luminosity, rendering a relatively weak outflow compared to typical expectation for effective feedback.
%which is about 0.001\% of the bolometric luminosity in 3C 191. 
%In a companion paper (Sharma et al, in preparation), we analyze the absorption data of the outflows based on observations with VLT/Xshooter.  We find a good agreement between the physical parameters of the outflow (distance from the central source, $\dot{M}$ and $\dot{E}_{\rm kin}$), deduced from the IFS and absorption analysis.

\end{abstract}

\keywords{galaxies: evolution---quasars: emission lines---quasars: supermassive black holes}


\section{Introduction} \label{sec:introduction}

In the picture of co-evolution between supermassive black holes residing in galaxy centers and their hosts, feedback from active galactic nuclei (AGN) may play a key role in the self-regulation of the galaxy growth. The AGN-driven large-scale outflows are known to be capable of clearing out the surrounding gas and quenching future star formation
(e.g., \citealt{Tabor1993,Silk1998,Loeb2004,DiMatteo2005,Springel2005,Ostriker2010,Novak2011,Soker2011,Choi2014,Nims2015,Ciotti2017,CostaSouza2024}). Roughly 20\% of quasars, show blueshifted broad absorption lines (BAL; \citealt{Weymann1991,Hewett2003,Reichard2003,Arav2008,Knigge2008,Scaringi2009,Allen2011}), indicating the presence of outflows. Theoretical modeling demonstrates that massive, sub-relativistic outflows may indeed produce efficient feedback on their host galaxies (e.g., \citealt{Ciotti2010,McCarthy2010,Ostriker2010,Soker2010,Soker2011,Faucher-Giguere2012, Choi2014,Mo2023}). However, direct observational evidence of black hole-galaxy self-regulation remains limited, particularly at high redshift. This is largely due to the limitations in spatial resolution and sensitivity of current instruments, both of which are crucial for characterizing AGN outflows.

In the cosmic history, the epoch around $z = 2 - 3$ is particularly important for investigating quasar feedback because it marks the peak of star formation and quasar activity in the universe \citep{Boyle1998,Hopkins2008}. Observing high-redshift AGN feedback is challenging, though an appreciable amount of investigations have been published (e.g. \citealt{Harrison2012,Genzel2014,Carniani2015,Cresci2015,Shen2016,Bischetti2017,Kakkad2020,Cresci2023,Veilleux2023,Vayner2024,Roy2024}). 

As yet, in several tens of AGN and quasars, the outflow has been observed via UV absorption lines by high-resolution spectrometry, such as VLT/UVES, VLT/X-Shooter and HST/COS (e.g., \citealt{Hamann2001,Moe2009,Aoki2011,Edmonds2011,Borguet2012,Chamberlain2015,Dunn2010,Borguet2012a,Arav2013,Borguet2013,Xu2020}).
The density of the absorber can be measured with the metastable excited-state absorption line. Combined with photoionization modeling, the extent, mass, and energetics of outflows are measurable. These results show that a large fraction of high-velocity outflowing gas is located at galactocentric distances up to tens of kpc (see \citealt{Arav2013,Arav2018} for a review), comparable to the spatial extent of the host galaxies. However, these results depend on multiple assumptions including the opening angle of the outflow, a quantity not measurable via absorption lines. In addition, absorption line analyses provide no information on the morphology and spatial structure of the outflow except along the line of sight direction.
 
In recent years, the ionized gas surrounding AGNs extending to kpc scales has been spatially resolved with Integral field spectroscopy (IFS) observations (e.g. \citealt{Nesvadba2008,Barbosa2009,Rupke2013,Shih2014,Liu2013a,Liu2013,Liu2014,Liu2015,Davies2015,McElroy2015,Alexander2010a,Vayner2024,Roy2024}). 
This technology enables detection of outflows around both radio-loud (e.g. \citealt{Nesvadba2006,Nesvadba2008,Husemann2019,Vayner2021,Ulivi2024}) and radio-quiet quasars (e.g. \citealt{Liu2013,Liu2013a}) across low (e.g. \citealt{Rupke2013a,Harrison2014}) and high redshift (e.g. \citealt{Alexander2010,CanoDiaz2012,Alexandroff2013,Kakkad2020,Vayner2024,Roy2024}).
%This technology enables detection of outflows both at low redshift $z < 0.5$ (e.g., \citealt{Rupke2013a,Harrison2014}), intermediate redshift (e.g. \citealt{Liu2013,Liu2013a,Liu2014,Shen2023}) and high redshift (e.g., \citealt{Nesvadba2006,Alexander2010,CanoDiaz2012,Alexandroff2013,Kakkad2020,Vayner2024,Roy2024}).
These observations provide evidence for outflow features extending up to galactic scales. %ubiquitous galaxy-wide gas moving with velocities of hundreds of km s$^{-1}$. 

IFS has the obvious advantage over absorption line analysis as it directly measures the spatial extent of the outflows with emission-line gas. However, the latter provides more abundant spectroscopic information covering a wide range of ionization and density in a single spectrum or in a few spectra. It enables the probe of the physical conditions in the ionized outflowing gas \citep{Arav2013}. Therefore, the IFS mapping of AGN outflows directly tests the outflow locations derived from UV absorption lines, which provides strong constraints on the physical assumptions and parameters of the outflow model if the absorption and emission line features are associated with the same wind (e.g. \citealt{Liu2015}).
The combination of UV mini-BAL analyses and IFS observation may shed light on the morphology and kinematics of outflows and probe the physical conditions in the outflowing gas.

3C 191 is a compact steep-spectrum, radio-loud quasar (with a jet power of 10$^{47.06}$ erg s$^{-1}$, \citet{Corbin1991}), with a central black hole of 10$^{9}$ M$_\odot$ \citep{Liu2006}. It shows a bipolar, lobe-dominated linear radio structure along PA 10 deg with a total extent of $\sim$ 5 arcsec, and a jet-like morphology to the south \citep{Lonsdale1993,Akujor1994}.
There are radio hotspots with projected galactocentric distances of 41 kpc, from which an approximate age of the radio source of $3 \times 10^{6}$ yr can be inferred \citep{Fanti1995}.
Moreover, 3C 191 could be a double–double radio source whose jet has been interrupted and restarted. The HST image of 3C 191 only detects the central source, not an optical counterpart to the jet \citep{Sambruna2004}.
The optical spectral index, $\alpha_{opt} = 0.7$ \citep{Barthel1990}, quite similar to the typical radio quasar value of $\alpha_{opt} = 0.5$ \citep{Brotherton2001}, indicates that there is little reddening along our line of sight to 3C 191 \citep{Willott2002}.
Extended X-ray emission around 3C 191 has been detected by Chandra observations \citep{Sambruna2004,Erlund2006}.
The extended X-ray emission is preferentially aligned along the radio jet direction. 
This X-ray emission is due to inverse-Compton scattering of the cosmic microwave background (CMB) photos with electrons in the jets and lobes.

3C 191 also shows UV absorption troughs and is identified as a mini-BAL quasar, which is defined as having absorption troughs with a full width at half maximum between 500 and 2000 km s$^{-1}$ (e.g. \citealt{Maiolino2024}). High-resolution spectroscopy has shown that the absorbing gas is blueshifted with velocities between 0 and -1000 km s$^{-1}$ \citep{Hamann2001,Sharma}.
The UV absorption lines appear to arise from a region of the galaxy-wide ($\approx$ 10 kpc) radiative outflow interacting with the disturbed interstellar medium (ISM) in the elliptical host galaxy \citep{Williams1975}. 
3C 191 has a large integrated rotation measure of $\sim -131$ to 1700 rad m$^{-2}$ along the jet, which is unlikely to be intrinsic to the jet but rather probably arises in the absorbing system of associated absorption lines (AAL) produced by a thin shell at $z_{abs}=1.9453$ \citep{Kronberg1990,Perry1990}. The residual rotation measure is two orders of magnitude larger than other quasars with comparable radio morphology and is consistent with a thin shell of 25 kpc across. The radial component of the magnetic field in the absorber is of order 0.4-4$\mu$G \citep{Kronberg1990}.

We conduct an IFS mapping of the mini-BAL quasar 3C 191 at $z \sim 2$, using the Spectrograph for Integral Field Observations in the Near Infrared (SINFONI) aided by adaptive optics (AO) and equipped on the Very Large Telescope (VLT).
This paper is structured as follows: After an overview of 3C 191 in previous studies, we describe observations and data reduction in
Section 2. In Section 3, we present the analysis of the spectral data and measure the gas kinematics. In Section 4, we discuss the origin of the ionized gas and the possible geometry of the outflow. In Section 5, we discuss the outflow kinematics and the relation between emission-line and absorption-line outflows and the origin of the \oiii\ outflow.
We conclude with a summary in Section 6. Throughout this paper, we adopt a cosmology with $H_{0}$=0.7 km s$^{-1}$ Mpc$^{-1}$, $\Omega_{m} = 0.3$, $\Omega_{\Lambda} = 0.7$.

\section{Observations and Data Reduction} \label{sec:style}

\subsection{Observational Strategy and Data reduction}  \label{sec:dr}

The target of this work, the quasar 3C 191, was observed by VLT using the near-IR integral field SINFONI with AO in December 2017 and January and March 2018 (program ID: 097.B-0570(B), PI: Benn).

We conduct observations in H-band ($\lambda$ $\sim$ 1.45 - 1.85 $\mu m$), covering the rest-frame 4915 $\mathrm{\AA}$ to 6271 $\mathrm{\AA}$ for an object with $z \sim 2$. 
The field of view (FOV) covers an average of $3'' \times 3''$ with a angular resolution of $0.24'' \times 0.24''$, roughly corresponding to 2 kpc at the average redshifts of $z = 2$ and a medium spectral resolution of R = 3000. 
The typical seeing is 0.4 - 0.7 arcsec. A standard star was observed for telluric correction. 
The total on-source integration time is  16 $\times$ 300 sec in four nights.
The airmass lies between $\sim$ 1.2 and $\sim$ 1.5.

%\subsection{Data reduction}  \label{sec:dr}

After removing cosmic rays from the raw data using the L.A. Cosmic procedure \citep{Dokkum2001}, we reduce the raw data using the ESO-SINFONI pipeline. The final data cubes have a spatial scale of $0.05\arcsec \times 0.05\arcsec$ with the ESO-SINFONI pipeline. We combine these four cubes to construct a median datacube.
%We combine these four cubes with medium value in each spatial pixel (spaxel) to produce the final science data cube. 
The point-spread function (PSF) is estimated by fitting a circularly symmetric 2D Gaussian profile to acquisition star exposures, which results in a full width at half-maximum (FWHM). 
%The PSF FWHMs correspond to the effective resolution for our target.

\subsection{Flux Calibration}  \label{sec:dr}
Because the absolute flux of our observation block was not calibrated, we flux-calibrate our data using the spectra of 3C 191 from VLT/X-shooter (program ID: 092.B-0393(B), PI: Leipski). The spectra are collected by a 0.9$\arcsec$-wide slit with the position angle of 0$^{\circ}$ (north to south) at a seeing of 1.47$\arcsec$.

We choose the \oiii\ $\lambda$5007 to perform the flux calibration. To simulate the VLT/X-shooter slit observations, we convolve the IFS image at each wavelength with a Gaussian kernel whose FWHM satisfies $\rm FWHM^2 + {0.24''}^2 = {1.47''}^2$ to mimic the VLT/X-shooter observing conditions, and then we extract the spectrum using a 0.9$\arcsec$-wide slit. We then measure the flux of \oiii\ $\lambda$5007 and compare the flux to that of the VLT/X-shooter. Finally, we multiply by a factor to normalize the SINFONI spectrum to match the X-shooter spectrum. Therefore, the \oiii\ absolute flux of the calibrated SINFONI spectrum is equal to that of the X-shooter spectrum.

We test our flux calibration using the spectra of 3C 191 from CGS4 spectrograph on UKIRT \citep{Wilman2000}.  The spectra are collected by a 1.22$\arcsec$-wide slit with a position angle of 0$^{\circ}$ (north to south) at a seeing of 1.3$\arcsec$. We perform flux calibration using the same method as described above. The calibrated \oiii\ luminosity from the UKIRT/CGS4 spectra is about twice as high as that from VLT/X-shooter. Both the calibrated \oiii\ luminosity from UKIRT/CGS4 and VLT/X-shooter are consistent within the uncertainty. This consistency indicates that our flux calibration is reliable. Given that the X-shooter spectrum has significantly better quality than the CGS4 spectrum (with signal-to-noise ratios of 62 versus 7), we rely on the X-shooter spectrum for flux calibration.

\section{IFS data analysis} \label{sec:dr}

\subsection{systemic redshift based on \oiii}  \label{sec:redshfit}

\begin{table}[htb]\footnotesize
\caption{Redshift measurements of 3C 191.}
\hspace{-0.6in}
\begin{tabular}{ccc}
\hline
\hline
 $z$ & emission line &  reference  \\
\hline
1.9523          & \civ, \heii\ and \ciii & \citealt{Burbidge1966} \\
1.955                & \lya, \civ\ and \heii & \citealt{Stockton1966} \\
1.9543       &  \lya  &  \citealt{Bahcall1967} \\
1.9587      &  \civ  &  \citealt{Bahcall1967} \\
1.956          & \civ, \heii, \oiiij\ and \ciii  & \citealt{Tytler1992} \\
1.9682        & ensemble of the emission lines &  \citealt{AdelmanMcCarthy2008} \\
1.9527$\pm0.0003$     & \oiii\ & this work \\  
\hline
\end{tabular}
\hspace{0.6in}
\label{tab:red}
\end{table}

The systemic redshift is crucial for investigating the kinetics of the gas.
In previous works, the redshift of 3C 191 has been measured using different line tracers (e.g. \citealt{Burbidge1966,Stockton1966,Bahcall1967,Tytler1992}), and their results are listed in Table ~\ref{tab:red}.
Previous measurements were either made with UV emission lines or with spectra of lower quality or poor resolution. We note that previous work (e.g. \citealt{Burbidge1966,Stockton1966,Burbidge1967,Bahcall1967}) using BLR emission lines to determine the systemic redshift is relatively unreliable (with $ z_{err} \sim 0.0057$), as the broad-line region may not be virialized. This method may be affected by bulk motion, inflow/outflow, and the central black hole may not be stationary relative to the galaxy (e.g. \citealt{Vietri2020}). Therefore, we re-measure the redshift.
The systemic redshift of 3C 191 can not be accurately determined from the stellar absorption lines as the stellar continuum of the host galaxies is too faint to be detected reliably in our data. Hence, we use emission lines from the narrow-line region, instead.

We use the narrowest component of the \oiii5007 emission in the galaxy center.
We extract the spectrum by applying an aperture with a size of 0.25$\arcsec$ centering on the quasar nucleus  which is the peak-flux location of the PSF for 3C 191, as shown in Figure ~\ref{fig:example} (black lines). The continuum and the \feii\ of this spectrum is subtracted (detailed in Section~\ref{sec:fitting}). The individual components of the best-fit spectra are depicted by orange dashed lines in the same figure. The \oiii\ $\lambda$ 5007 line is fitted with a combination of 3 Gaussians, assuming that the central component is the narrowest. 
We determine the systemic redshift to be $z_{em} = 1.9527$ $\pm$ 0.0003 from the peak wavelength of this central narrowest component ($\sigma \sim$ 120 km s$^{-1}$).

An independent redshift can also be measured using the \oii\ doublet observed by VLT/X-shooter. We measured a $z_{em} = 1.9527$ $\pm$ 0.0003 using \oii $\lambda\lambda$3727,3729 doublet (Figure ~\ref{fig:example}). \oii-based redshift is in good agreement with the above \oiii\ result. In view of the blending of the \oii\ doublet, we adopt the above \oiii-based systemic redshift. 

\begin{figure}[htb]
\centering
\includegraphics[width=0.45\textwidth]{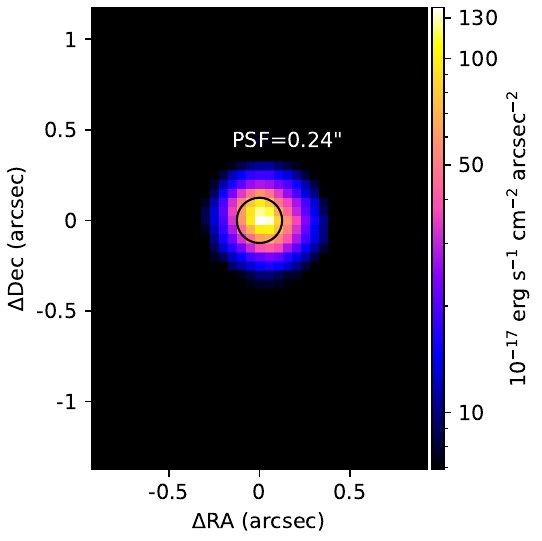}
\caption{The PSF model is integrated over the rest-frame wavelength range of 5030 $-$ 5800$\AA$. The black open circle depictes the FWHM of the PSF (0.24$\arcsec$).
\label{fig:psf}}
\end{figure}

\begin{figure*}[htb]

\includegraphics[width=0.44\textwidth]{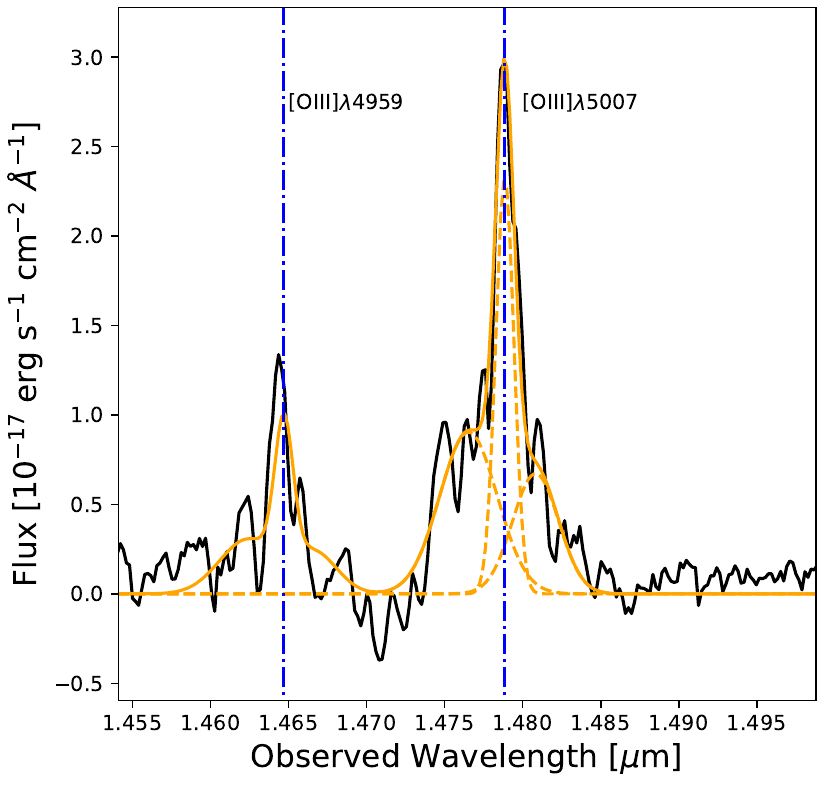}
\includegraphics[width=0.457\textwidth]{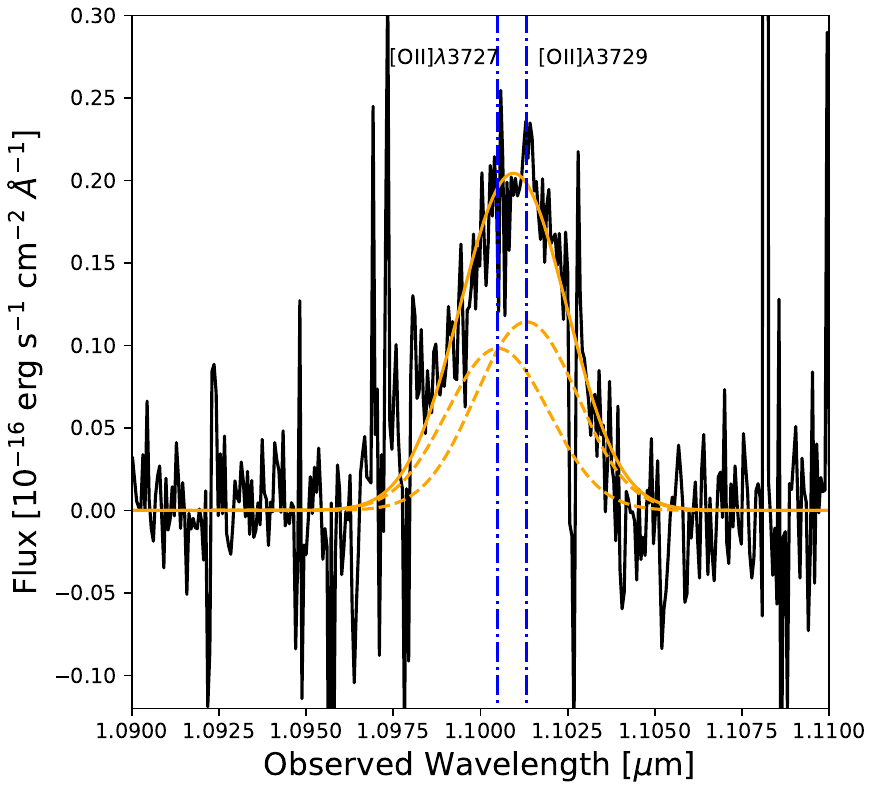}
\caption{Line fitting of the VLT/SINFONI (left) and VLT/X-shooter (right) data, for the purpose of the redshift determine. Left panel: The black solid line represents the observed \oiii\ spectrum from VLT/SINFONI, summed within a 0.25$\arcsec$ aperture centered on the quasar location, with the continuum and \feii\ emission subtracted. The orange solid line shows the total fit, while the individual components are represented by orange dashed lines. The vertical blue line indicates the center wavelength of the narrowest component, which is used to determine the systematic redshift. Right panel: The black solid line represents the observed \oii\ spectrum from VLT/X-shooter. The systematic redshift is determined by fitting the \oii $\lambda\lambda$3727,3729 doublet, with the orange solid line showing the total fit and the orange dashed lines showing the individual components.} 
\label{fig:example}
\end{figure*}

\subsection{Multi-Gaussian fits}  \label{sec:fitting}

To search for extended \oiii\ emission around 3C 191, we first subtract the continuum and \feii\ emission from the data cube. We extract the spectrum in each spaxel, fit the spectrum with a model combining a power law and \feii\ in the wavelength interval on both sides of the \oiii\ vicinity, 4918-4930$\AA$, and 5030-5900$\AA$, where the continuum is free of any line emission. Note that \oiii\ lines lie on the blue part of the waveband, and \hb\ is not covered. We use the \feii\ template from \citet{Tsuzuki2006}, convolving it with a Gaussian profile with a velocity dispersion $\sigma \sim$ 2000 km s$^{-1}$ and then subtract the \feii\ emission.
We obtained a \oiii\ cube by subtracting the continuum and \feii\ emission from the spectrum in each spaxel. The initial spectrum prior to continuum and \feii\ subtraction is shown in Appendix.

\begin{figure*}[htb]
\centering
\includegraphics[width=0.45\textwidth]{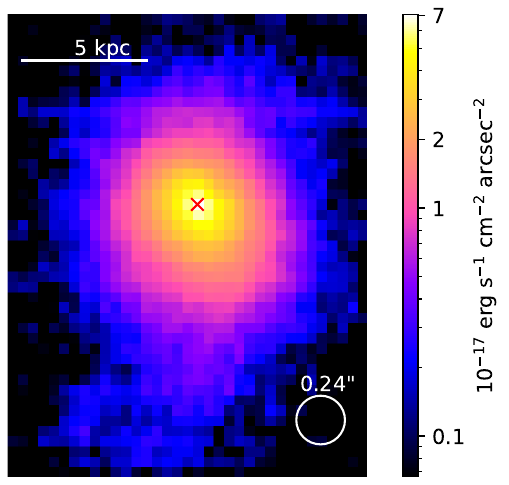}
\includegraphics[width=0.45\textwidth]{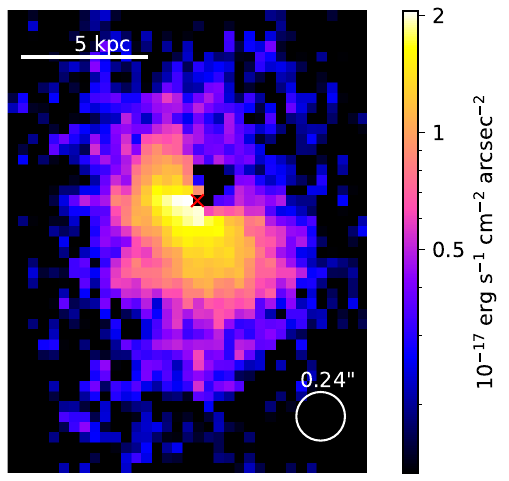}
\caption{Integrated flux intensity of \oiii\ (erg s$^{-1}$ cm$^{-2}$ arcsec$^{-2}$) from collapsing the IFS datacube directly after subtracting continuum and the \oiii\ map after subtracting the PSF. PSF is depicted by the open circle on each map. The red cross marks the QSO position which is the peak-flux location of the PSF for 3C 191. These \oiii\ maps reach a sensitivity of $\sim 2 \times 10^{-19} {\rm erg\ s^{-1}\ cm^{-2}\ arcsec^{-2}}$ on the backgound.
\label{fig:subpsf}}
\end{figure*}

We remove a point source at the quasar position to help reveal the extended emission. We construct another PSF model by integrating the rest-frame wavelength interval of 5030-5800$\AA$, which is free of any line emission and artifacts. This PSF model, shown in Figure ~\ref{fig:psf}, has an FWHM consistent with the one derived from the acquisition stars. For each frame of the datacube, we scale the PSF model to match the central peak brightness and then subtract it. Figure ~\ref{fig:subpsf} shows the integrated \oiii\ $\lambda$5007 intensity map by collapsing the \oiii\ cube after subtracting the PSF in the wavelength range of 4985-5025$\AA$. Both maps show only where the signal exceeds 3$\sigma$ ($\sigma$ is the standard deviation of background) against the background. Our observation reaches a sensitivity of $6 \times 10^{-19}$ erg s$^{-1}$ cm$^{-2}$ arcsec$^{-2}$.
The \oiii\ emission, after the PSF removal, shows an extended structure. This extended \oiii\ emission surrounds the quasar with a projected radii of $R =$ 5 to 11 kpc, extending further to the south.
The morphology of \oiii\ emission after PSF subtraction roughly shows a biconical structure.

\begin{figure*}[htb]
\centering
\includegraphics[width=0.24\textwidth]{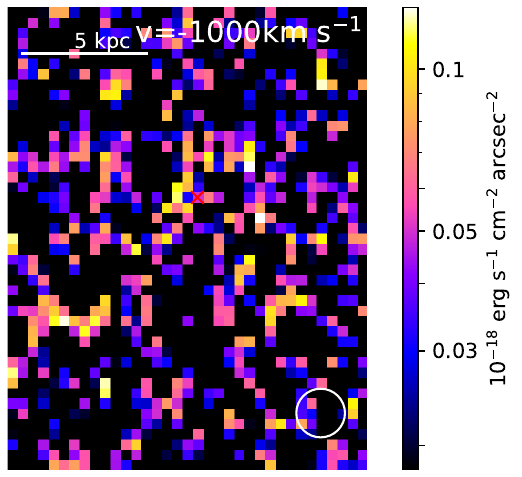}
\includegraphics[width=0.24\textwidth]{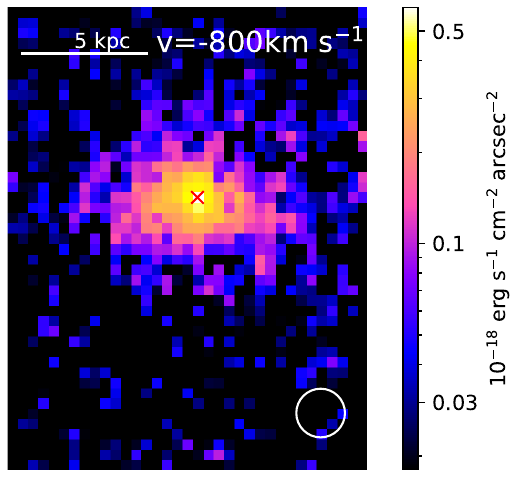}
\includegraphics[width=0.24\textwidth]{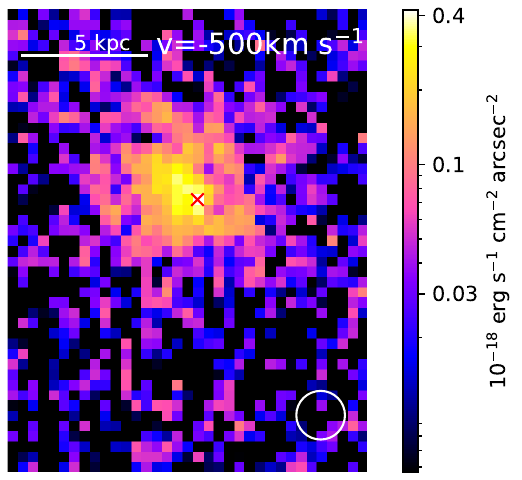}
\includegraphics[width=0.24\textwidth]{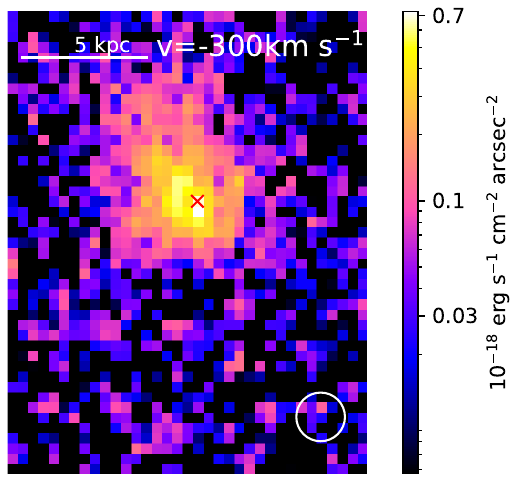}
\includegraphics[width=0.24\textwidth]{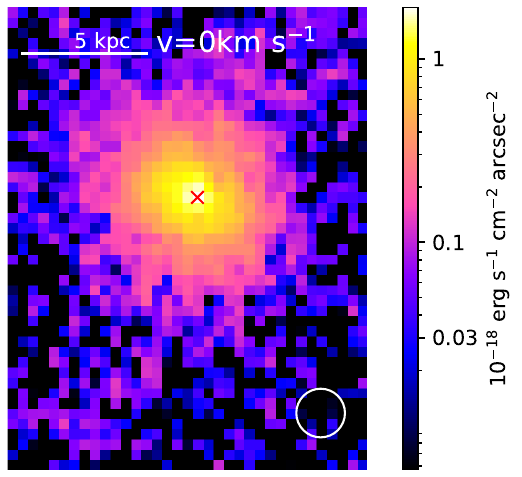}
\includegraphics[width=0.24\textwidth]{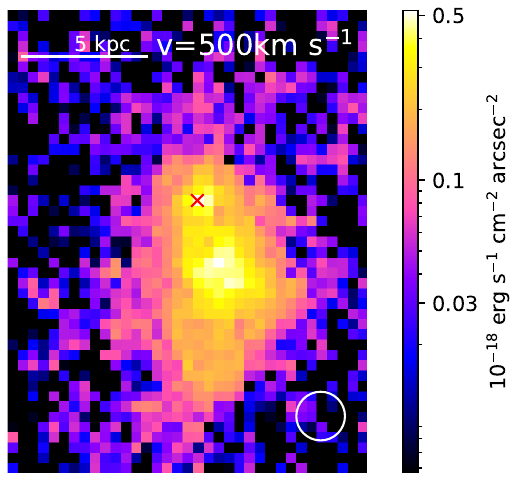}
\includegraphics[width=0.24\textwidth]{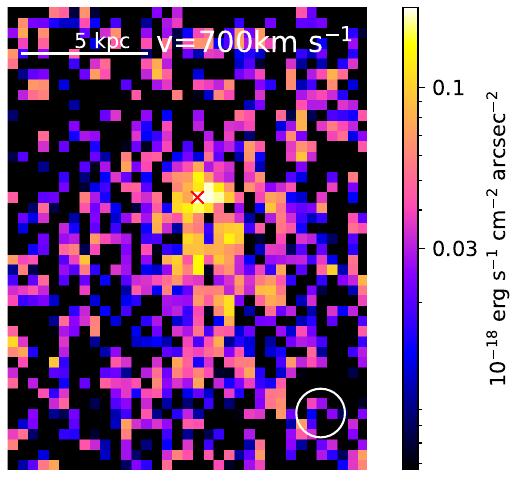}
\includegraphics[width=0.24\textwidth]{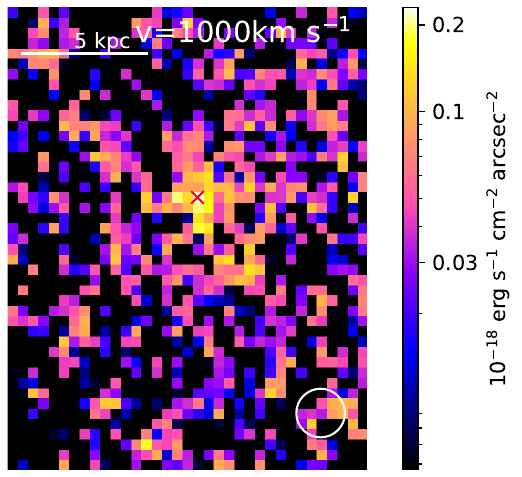}
\caption{Spatial distribution of \oiii\ at different velocity. PSF is depicted by the open circle on each map. The red/black cross marks the QSO position.
\label{fig:im}}
\end{figure*}

The eight panels in Figure ~\ref{fig:im} depict the spatial distribution of \oiii\ at different velocities. Two \oiii\ blobs are found to have a velocity of 500 km s$^{-1}$, of which one is located in the center, while the other is in the southwest.
The unshifted \oiii\ component is located in the center, while the blueshifted and redshifted components are north-ward and south-ward, respectively.
Consequently, we fit the \oiii\ profile across all positions with three velocity components: blue-shifted, red-shifted, and unshifted (i.e. systemic).

In most positions of the field of view, the profile of \oiii\ $\lambda$ 5007 $\mathrm{\AA}$ emission line appears complex. This may be due to the 3-dimensional velocities of the multiple moving gas components projected onto the line of sight.

We analyze the \oiii\ cube to delineate the gas kinematics.      
We assume the \oiii\ doublet originating from the same upper level, fit the doublet imposing the same central velocity and velocity dispersion, and the intensity ratio $I(5007)/I(4959) \sim 3$.  
\oiii\ gas clearly shows three components with different velocities, indicating blue-shifted, red-shifted components, and an unshifted component presumably in bounded motion in the quasar host galaxy (Figure ~\ref{fig:im}). 
These three components spatially overlap with each other (Figure ~\ref{fig:subpsf}). A single Gaussian function is inadequate to reproduce the asymmetric velocity profiles in each spatial element (spaxel), so we fit a combination of multiple Gaussians to the \oiii\ $\lambda\lambda$4959,5007 doublet by minimizing $\chi^2$ using the Python package MPFIT following \citet{Liu2013}. We experiment with single, double, and triple Gaussians, finding that no more than 3 Gaussians are needed to fit the \oiii\ profiles, as found by \citet{Liu2013}. When fitting to a triple Gaussian model, we set the initial values of the central velocity and velocity dispersion as the parameters that fit the nucleus spectrum (Figure ~\ref{fig:example}). The parameters of all models are not subject to any range restrictions. For each model, we calculated the Reduced chi-square statistic between data and models and the corresponding probability that the model fits the data. We demand that if the probabilities of two models differed by a threshold of 0.01, which is a widely used fiducial criterion of significance level for hypothesis testing in statistics, the model with fewer Gaussians is adopted. This method has been widely used in previous similar work (e.g. \citealt{Liu2013,Liu2014,Liu2015}).

\subsection{Outflow Parameters}  \label{sec:dr}

We describe the emission line profiles with the following parameters \citep{Whittle1985,Liu2013}.

(i) integrated intensity of the \oiii\  $\lambda$ 5007 $\mathrm{\AA}$ emission line.

(ii) the median velocity $V_{med}$.

(iii) the line width $W_{80}$, calculated as the width in terms of velocity that encloses 80\% of the total flux: $W_{80} = v_{90} - v_{10}$. For a single Gaussian profile, $W_{80} = 2.563 \times \sigma$, close to the conventionally used FWHM.

\begin{figure*}[htb]
\centering
\includegraphics[width=0.30\textwidth]{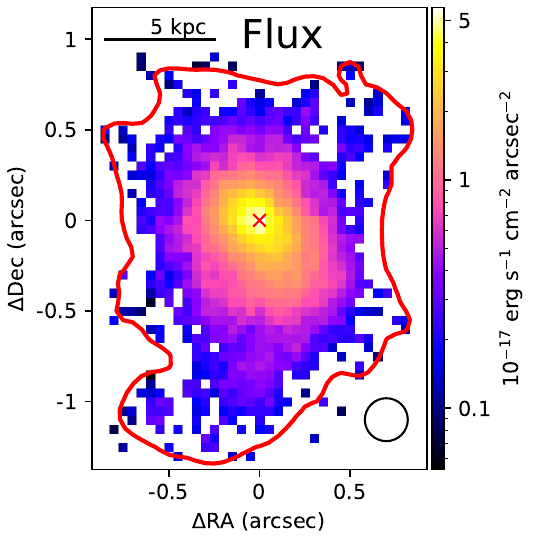}
\includegraphics[width=0.31\textwidth]{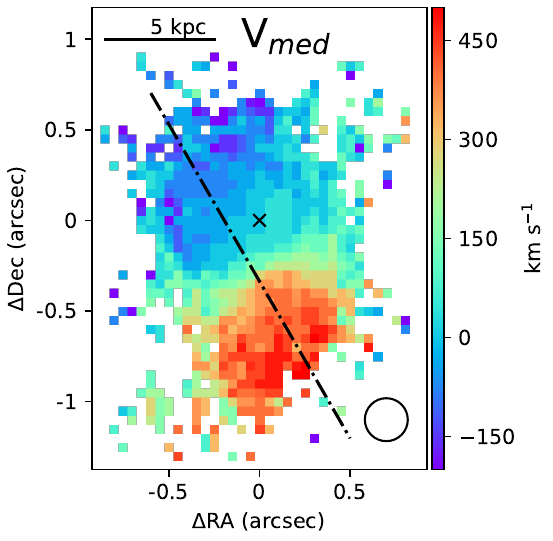}
\includegraphics[width=0.31\textwidth]{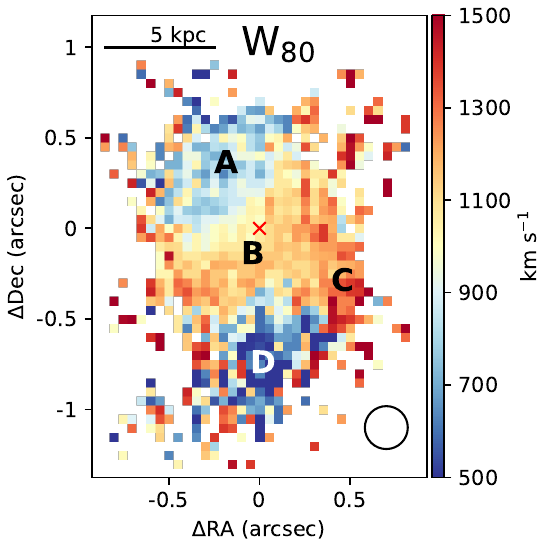}
\caption{Non-parametric measurement of \oiii\ ionized gas around 3C 191. From left to right are: flux intensity of \oiii\ (erg s$^{-1}$ cm$^{-2}$ arcsec$^{-2}$), median velocity ( km s$^{-1}$), line width ($W_{80}$, km s$^{-1}$). Only spaxels where the peak of the \oiii\ $\lambda$5007 line is detected with S/N $>$ 3 are plotted. PSF is depicted by the open circle on each map. The red/black cross marks the QSO position. The red contours trace the morphology of \oiii\ ionized gas. We use dashed line to represent the axes of the outflow by eye.
\label{fig:total}}
\end{figure*}

\begin{figure*}[htb]
\centering
\includegraphics[width=0.9\textwidth]{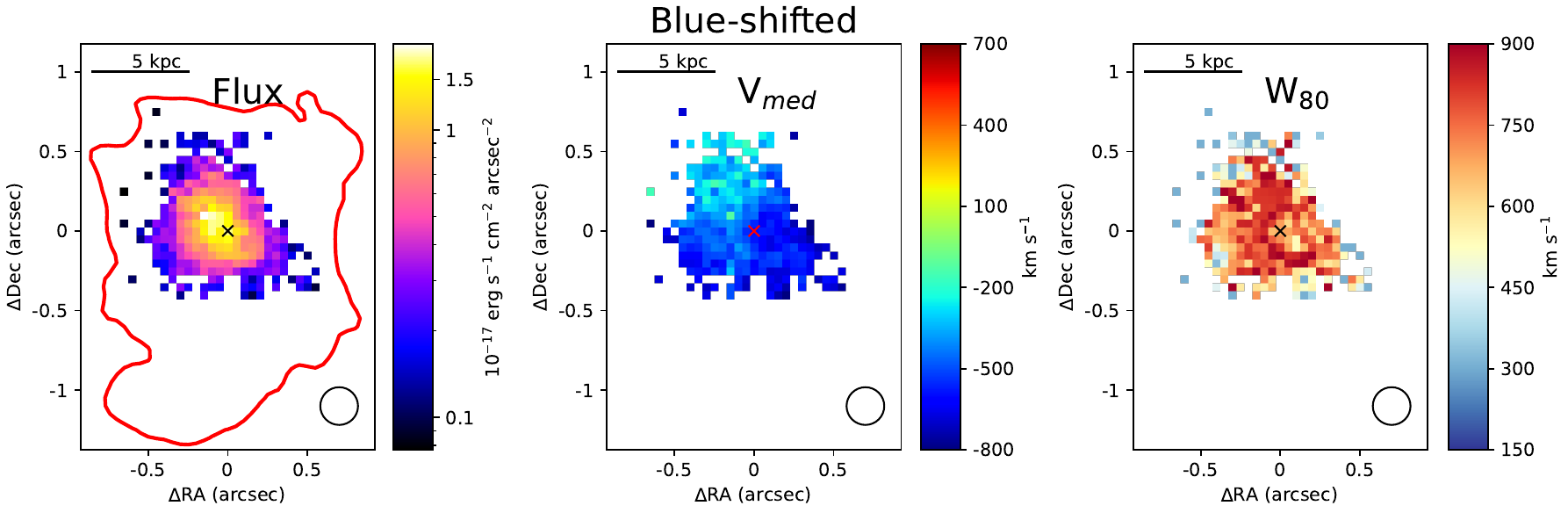}
\includegraphics[width=0.9\textwidth]{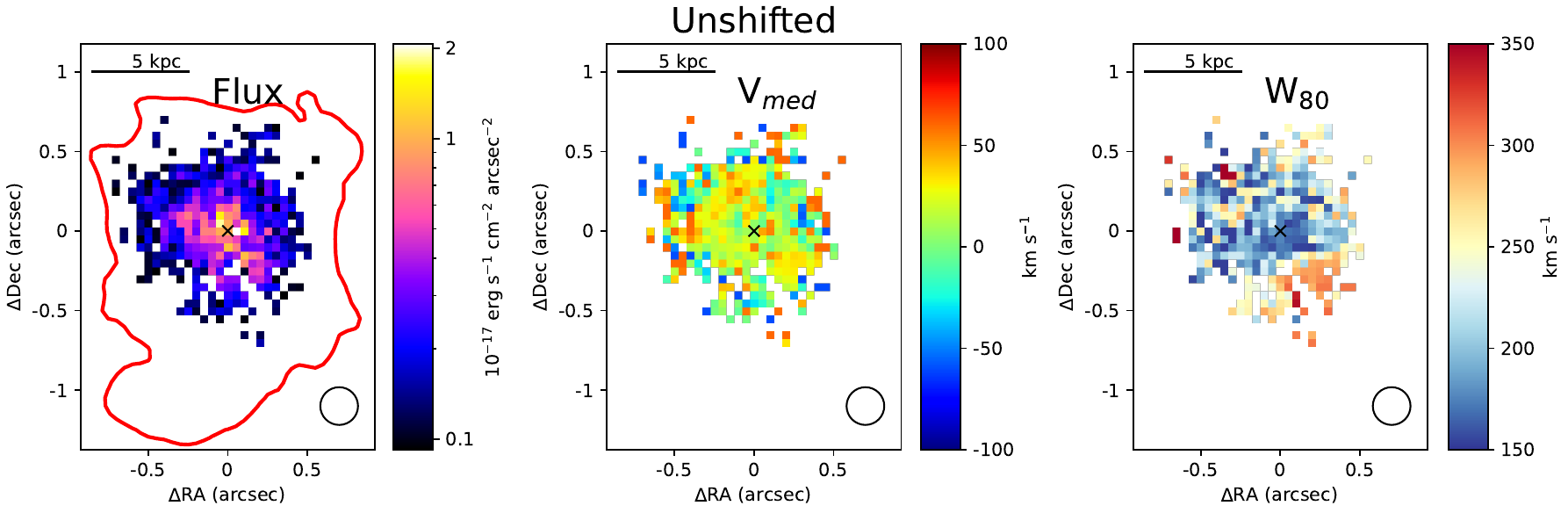}
\includegraphics[width=0.9\textwidth]{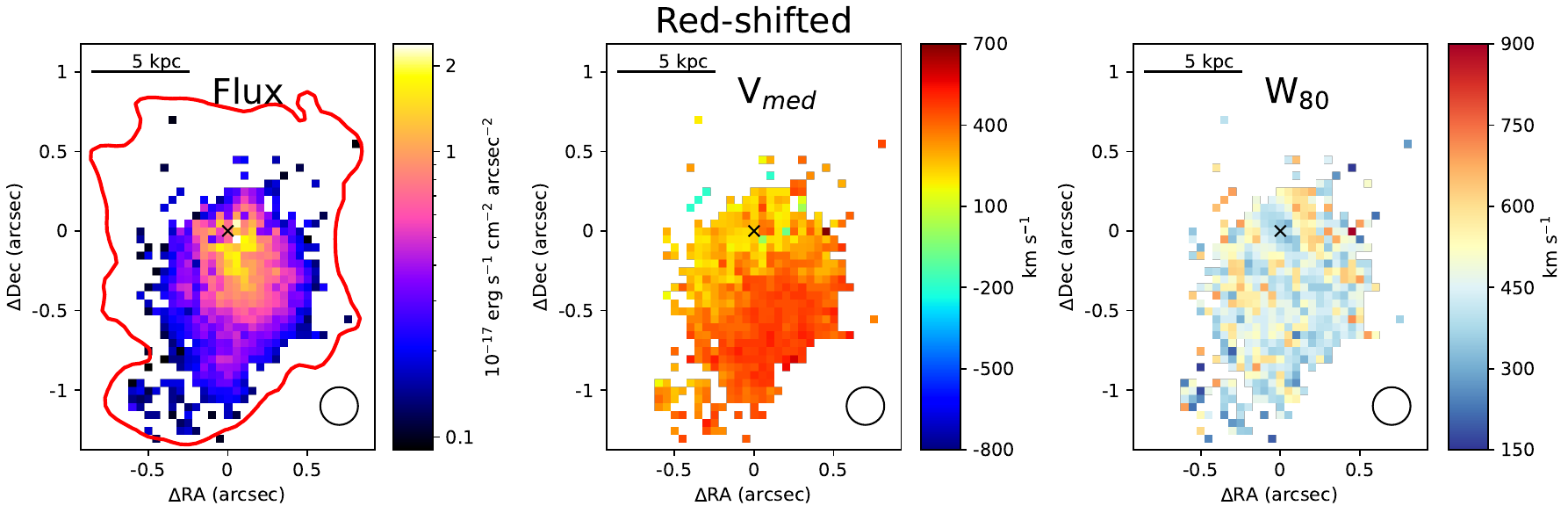}
\caption{The maps of the ionized gas for 3C 191. The three columns from left to right, the three rows from top to bottom are: \oiii\ $\lambda$ 5007 flux intensity (erg s$^{-1}$ cm$^{-2}$ arcsec$^{-2}$), median velocity (km s$^{-1}$), and velocity dispersion map ($W_{80}$, km s$^{-1}$) of blueshifted, unshifted and redshifted components of ionized gas with a cut-off at a S/N of 3, respectively. PSF is depicted by the open circle on each map. The red/black cross marks the QSO position. The red contours trace the morphology of \oiii\ ionized gas from Figure ~\ref{fig:total}.
\label{fig:fk}}
\end{figure*}

The advantage of nonparametric measurement is that they do not depend on multi-Gaussian decomposition, which may not be reliable. Figure ~\ref{fig:total} shows the parameter maps of the emission line as a whole, and Figure ~\ref{fig:fk} shows the parameter maps of the three components. The maps are created by keeping only those spatial pixels with a S/N ratio no less than 3.

\begin{figure}[htb]
\centering
\includegraphics[width=0.45\textwidth]{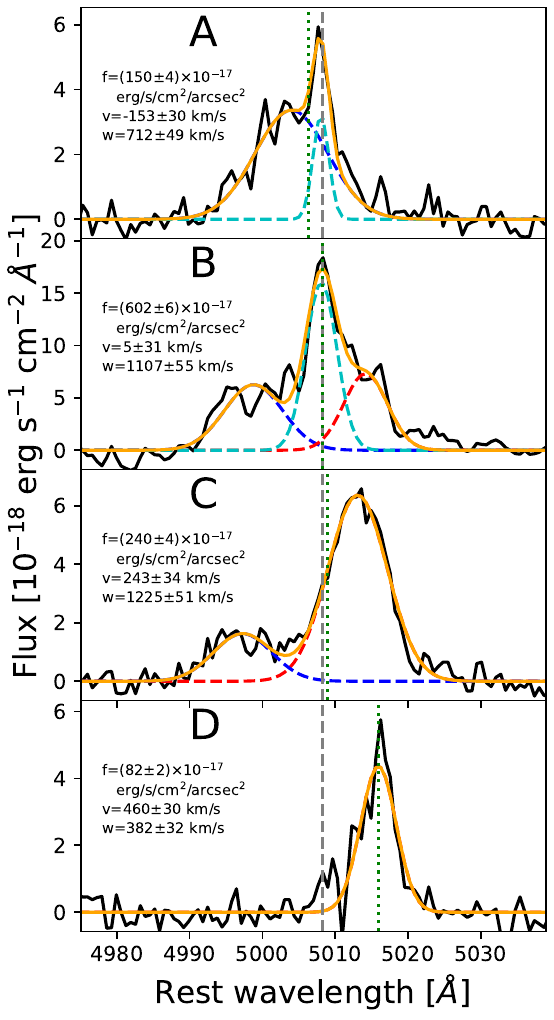}
\caption{We select four spatial positions to present the \oiii\ velocity profile therein, including central (``B''), redshifted (``C'', ``D'') and blueshifted (``A'') regions.  Locations of these regions are labelled by their IDs in Figure ~\ref{fig:total}. The fitted line is in orange, and the central narrow unshifted component and the blue-/redshifted components are in cyan, blue, and red, respectively. The median and zero velocity are marked by green dotted and grey dashed lines, respectively. The integrated intensity of the \oiii, $V_{med}$ and $W_{80}$ with their uncertainties for the individual regions are shown.
\label{fig:spec}}
\end{figure}

\begin{figure}[htb]
\centering
\includegraphics[width=0.45\textwidth]{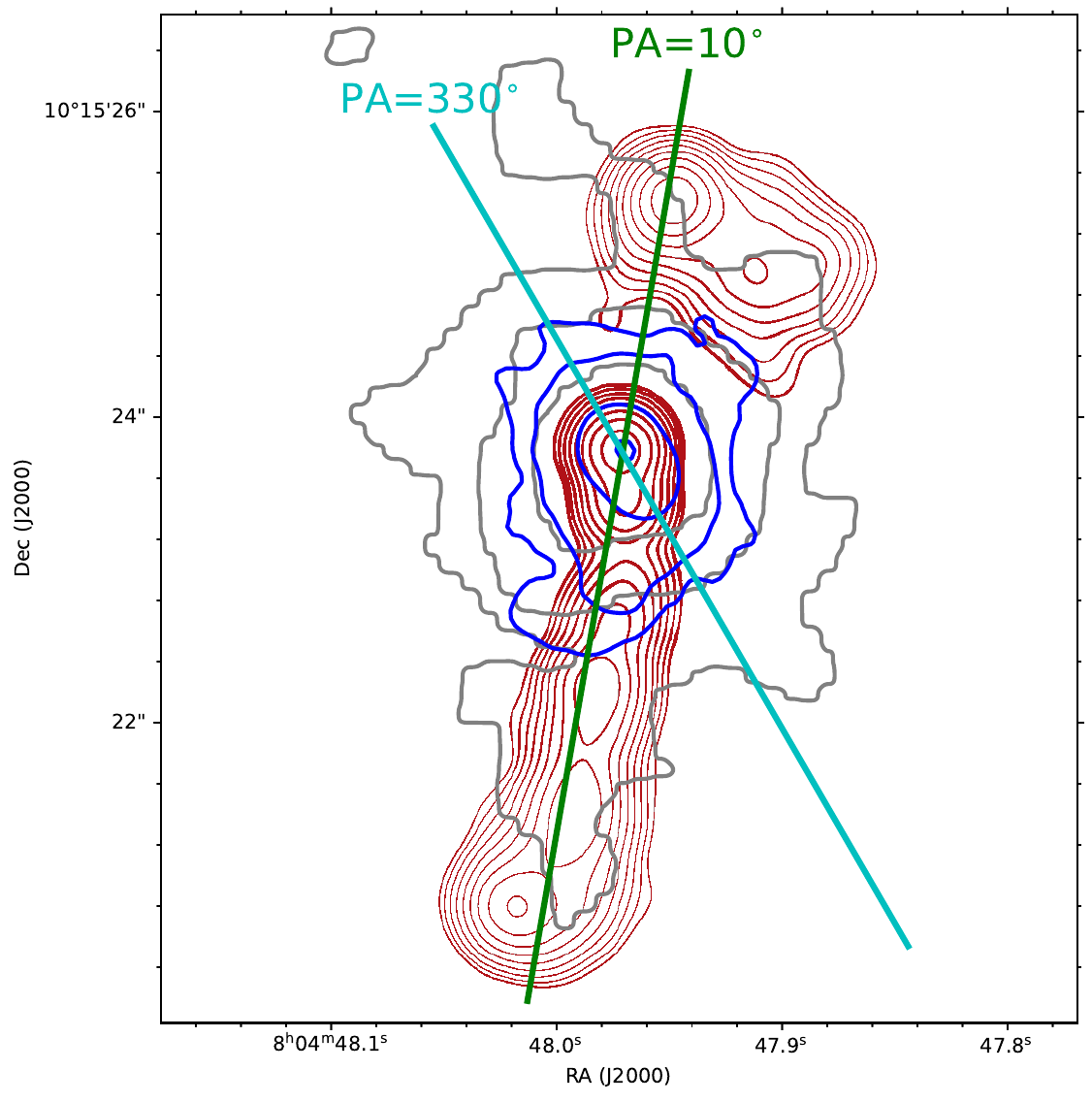}
\caption{The grey contours are Chandra ACIS-S observations of the radio jets of 3C 191 in the 0.5–8 keV energy range \citep{Sambruna2004,Erlund2006}, red contours are VLA observations from the quasar jet and lobes \citep{Kronberg1990}, while the blue contours representing surface brightness of the integrated \oiii\ with VLT/SINFONI observations in our data. The green and cyan soild lines represent the axes of the jet and outflow, respectively.
\label{fig:png}}
\end{figure}

Motivated by the different kinematic components apparent in the velocity and velocity dispersion, we extract and fit the \oiii\ lines in 4 positions with a size of 0.25$\arcsec$ $\times$ 0.25$\arcsec$ (referred to as ``A-D'') labelled in Figure ~\ref{fig:total}, and show them in Figure ~\ref{fig:spec}. ``B'' is the quasar position, ``A'' and ``D'' represent the regions with lower velocity dispersion, and ``C'' represents the region with higher velocity dispersion.
The results of the decomposition process described above are visually reasonable, implying that our decomposition is reliable and that the three components are naturally separated.

\section{The nature of the extended emission line feature} % (fold)
\label{sec:section_name}

The forbidden emission line \oiii\ $\lambda\lambda$4959,5007 doublet is a good tracer of ionized outflows on large scales. 
The \oiii\ ionized gas surrounding 3C 191 is spatially resolved by our IFS observation.
In this section, we describe the results of our analysis and discuss the nature of the extended emission line features, which are most naturally explained by outflows.

\subsection{Possibilities of tidal features or companion galaxies}  \label{sec:velocity}

The tidal debris and nearby small companion galaxies that are illuminated by the quasar \citep{Liu2009,VillarMartin2010,Perna2023} can produce extended narrow-line emission. However, these features are not proof of quasar feedback. We discriminate feedback from other possible origins of ionized gas based on the morphological correlation between the extended emission line region and the radio jet. The \oiii\ emission that surround radio-loud quasars with apparently disordered velocity fields, and little, morphological correlation with either the host galaxy or the radio structure, are often interpreted using a model of illuminated or shocked tidal debris \citep{Fu2009}. The \oiii\ emission around radio-loud quasars with well-organized velocity fields and morphological correlation with radio jets are likely related to outflows \citep{Ulivi2024}. In addition, $W_{80}$ lies in a rough range of 500-1500 km s$^{-1}$, which is significantly higher than that of tidal features ($\sigma \sim 50-100$ km s$^{-1}$, e.g. \citealt{Fu2009}). The high-velocity difference $\Delta V \sim 700$ km s$^{-1}$ and the fact that unshifted \oiii\ emission shows no velocity gradient in combination lead us to exclude the possibility of a rotating disk.

\subsection{Biconical outflow}  \label{sec:sperbubble}

As mentioned in Section ~\ref{sec:redshfit}, the \oiii\ line shows unambiguous triple-peak profiles in the center, and at other positions. These three peaks are also well separated (Fig ~\ref{fig:spec}). The bluer peak always corresponds to a negative velocity ( V$_{med}$ $\sim$ -500 km s$^{-1}$), the redder one always shows a positive velocity (V$_{med}$ $\sim$ 500 km s$^{-1}$), and the central one always shows velocity close to zero (Fig ~\ref{fig:fk} middle column). The parameters of these components demonstrate spatial continuity from the nucleus to outer regions, indicating physical authenticity.
Therefore, our analysis shows that the \oiii\ emission can be well decomposed into three components, which is similar to \citet{Liu2015,Zhao2021,Shen2023}. Hereafter, we refer to these three components as the blueshifted, redshifted, and unshifted components. This approach results in the decomposition shown in Figure ~\ref{fig:fk}.

The unshifted gas shows no spatial velocity gradient as evidence of rotation or an outflow. It has minimal radial velocity ($\sim$ 0 km s$^{-1}$) and is much narrower ($\sim$ 160 km s$^{-1}$) than the blueshifted and redshifted components ($v \sim 500$ km s$^{-1}$, $W_{80} \sim$ 800/500 km s$^{-1}$), implying an origin different from the latter two. In addition, the brightness profile peaks exactly at the quasar position, and the size of $\sim$ 5 kpc is much smaller than that of redshifted \oiii\ gas ($\sim$ 11 kpc). These are well consistent with typical narrow line regions in quasars, which typically originated in undisturbed gas in the host galaxy illuminated by the quasar.

The surface brightness of \oiii\ emission in the blueshifted and redshifted gases show smooth morphology, peaking close to the quasar position with a offset of 0.18$\arcsec$ and 0.11$\arcsec$, respectively and declining steeply with distance. The velocity maps show a clear gradient, and the velocity dispersion of the blueshifted and redshifted gases ($W_{80} \sim$ 800/450 km s$^{-1}$) is significantly larger than that of the typical NLR values. 
These are all consistent with the interpretation that the blueshifted gas is outflowing gas moving towards the observer, while the redshifted gas outflowing away from us.
We find the maximum $W_{80}$ value of $\sim$1600 km s$^{-1}$, which is comparable to that of known quasar outflows (e.g. \citealt{Liu2013, Kubo2022}).

The organized velocity gradient across the target can be interpreted by bipolar outflows (e.g. \citealt{RodriguezZaurin2013,Perna2015}), where the northern side is closer to and moving towards us, while the southern side is further away and moving away from us. The velocity and velocity dispersion maps are similar to that of a biconical outflow in numerical simulation  (see Fig. 3 in \citealt{Bae2016}).
Consequently, the smooth morphology of the \oiii\ gases, the velocity, and the high velocity dispersions of the gas,  all suggest that the blueshifted and redshifted gases are outflowing ionized gas.

Gas in the nucleus (denoted ``B'' in Fig. 4) and in the west (region ``C'') has broader line width with $W_{80}$ exceeding 1000 km s$^{-1}$. Gas in the north (region ``A'') and south (``D'') show narrower line widths (Fig ~\ref{fig:total}).
The positions with the maximum widths (``C'') are not in the center. The energetic outflow emanating from the AGN extends into the intergalactic medium in an expanding galaxy-scale bubble-like structure, which is known as a superbubble. 
We interpret these as a possible feature of a pair of superbubbles, though admittedly the evidence is weaker than that given in \citep{Shen2023}. Also, this might be due to the fact that, for a biconical outflow, the blueshifted and redshifted sides of the cone are overlaid in the vicinity of the center, meaning that both blueshifted and redshifted components contribute to the spectra, which broadens the observed line profile.

A pair of superbubbles is presumably symmetric about the central AGN (e.g. \citealt{Venturi2023}).
However, this is not the case in 3C 191.
In Figure ~\ref{fig:total}, we can find that the quasar is not located in the center of the velocity map of the total ionized gas. The position of AGN is approximately one PSF from the center of the velocity map, which is similar to the situation in  F2M1106 \citep{Shen2023}.
These asymmetries may be geometrically caused by the viewing angle, though selective dust obscuration or further complications are probably at play.

\section{Discussion} 

%\subsection{The systemic redshift}

%The systemic redshift is crucial for investigating the kinetics of the gas nebula. We measure the redshift of $z = 1.953$ using \oiii\ in the center of the quasar as well as using \oii, and the results are in good agreement. 

%In addition, the blueshifted and redshifted nebulae are shown symmetrical in morphology and dynamic, as expected from the picture of superbubbles \citep{Shen2023}. In detail, they lie along the north-to-south axis around the center ENLR region, and they are shifted in velocity with 420 km s$^{-1}$ and 430 km s$^{-1}$ about the systematic redshift, respectively.  Moreover, the redshift and average velocity of the extended \oiii\ emission are consistent with those measured in \citet{Wilman2000} ($z = 1.956$ and $v=170 \pm 230 km/s$) within the uncertainty. Note that they used slit spectroscopy which does not cover the full extended nebulae and may affect the velocity measurements. 

\subsection{Kinematic energy of the outflow} 

%\subsubsection{Kinematic energy and rate} 

In this subsection, we use the results from the spatially and kinematically resolved IFS spectra to measure the kinematic parameters of the outflows. We calculated the kinetic energy of the blueshifted and redshifted gas. Derived parameters are tabulated in Table ~\ref{tab:tab5}.

We measured the mass of ionized outflows using \oiii$\lambda$ 5007 (e.g \citealt{CanoDiaz2012,Liu2013,Harrison2014,Carniani2015}).
The mass of ionized gas can be estimated using \oiii\ luminosity $L_{\oiii}$ and electron density $n_{e}$ \citep{Osterbrock2006,Carniani2015} as:
\begin{equation}
 %\frac{M_{{\rm gas}}}{2.82\times10^{9}\,{\rm M_{\odot}}} = \left(\frac{L_{{\rm
%      H}\beta}}{10^{43}\,{\rm erg\,s}^{-1}}\right) \left(\frac{n_{e}}{100\,{\rm cm}^{-3}}\right)^{-1}
\tiny M = 0.8\times10^8 \rm M_{\odot} \left(\frac{C}{10^{[O/H]-[O/H]_{\odot}}}\right) \left(\frac{L_{[OIII]}}{10^{44} erg/s}\right) \left(\frac{<n_e>}{500 cm^{-3}}\right)^{-1},
\end{equation}
where $C = <n_e>^2/<n_e^2>$. We assume $C \approx 1$, [O/H]= [O/H]$_{\odot}$.
We assume that the absorption and emission lines trace the same outflow and therefore adopt an electron density of $n_e = 602 ^{+121}_{-124} \mathrm{cm}^{-3}$, derived from the column density ratio between the excited and resonance troughs of \siii\ AALs \citep{Sharma}. This value aligns with our measurement of $n_e = 272^{+340}_{-200} \mathrm{cm}^{-3}$ based on the \oii\ doublet \citep{Osterbrock2006} and is consistent with typical values for radio galaxies at $z = 2-3$ \citep{Nesvadba2008}.
Using the blueshifted and redshifted \oiii\ luminosities of 3C 191, which is from the integrated spectrum in the FoV, $L_{\oiii} = 5.3\pm0.6 \times 10^{43}$ erg s$^{-1}$, $L_{\oiii} = 4.7\pm0.5 \times 10^{43}$ erg s$^{-1}$,we find $M_{\rm gas} \sim 3.5\pm0.6 \times 10^{7}~ M_{\odot}$, $M_{\rm gas} \sim 3.1\pm0.3 \times 10^{7}~ M_{\odot}$, respectively.

We estimate the kinetic energy of the blueshifted and redshifted outflowing ionized gas to be:
\begin{equation}
E_{\rm kin}=\frac{1}{2} M_{\rm gas}v_{\rm gas}^2=3.3\pm0.8, 2.0\pm0.2 \times10^{56}~{\rm erg},
\end{equation}
where $v_{gas} = \rm v_{max}=|\Delta v| + 2\sigma$, (where $\rm \Delta v$ is the velocity shift between broad and core \oiii\ emission centroids, \citealt{Bischetti2017,Fiore2017}, here $\rm v_{med}$ of blueshifted and redshifted components represent $\rm \Delta v$), can be considered as representative of the bulk velocity of the outflowing gas.
We measure the average value of $v_{gas}$ of the blueshifted and redshifted from their velocity and velocity dispersion map (Figure ~\ref{fig:fk}) to represent the maximum shifted outflow velocity (see table~\ref{tab:tab5}). 
The blueshifted and redshifted gases have the velocity of $v \sim-975\pm46$ km s$^{-1}$, $v \sim+807\pm50$ km s$^{-1}$, and their projected distance to the nuclear is $R \sim 5^{+1.3}_{-0.2}, 11^{+0.5}_{-2}$ kpc, respectively. 
Then, we estimate a dynamical timescale of the blueshifted and redshifted gas to be $t_{dyn}$ = R/$v_{gas}$ $\sim 5.0\pm0.3 \times 10^{6}$, $1.3\pm0.1 \times 10^{7}$ yr, i.e., the time that the gas from the nucleus takes to reach such a distance with an average velocity of $v_{gas}$.
If we assume an expanding shell with constant density and velocity \citep{Ulivi2024}, the lifetime to be the dynamical timescale, we can also estimate the blueshifted abd redshifted gas mass flow rate to be $\dot{M}$ = M/$t_{dyn}$ $\sim$ 7.1-9.4, 2.4-4.0 M$_\odot$ yr$^{-1}$, and a kinetic energy rate $\dot{E}_{\rm kin} \sim$ 20.9-27.9, 4.8-8.6 $\times 10^{41}$ erg s$^{-1}$. We adopt $R \sim 5^{+1.3}_{-0.2}, 11^{+0.5}_{-2}$ kpc which is the outflow scale when SNR $\geq$ 3, at $R = $ 2.7 kpc which is the offest of the peak  of the redshifted \oiii\ with respect to the nucleus at $v \sim 500$ km s$^{-1}$ (Figure ~\ref{fig:im}) to estimate the range of $\dot{M}$ and $\dot{E}_{\rm kin}$.

We estimate the bolometric luminosity of the quasar by using the relation L$_{bol} \sim L(\lambda5100\AA)$ from \citet{Runnoe2012}. We obtain the 5100$\AA$ luminosity L(5100$\AA$)=1.0$\pm0.2$ $\times$ 10$^{42}$ erg s$^{-1}$ $\AA^{-1}$. The AGN bolometric luminosity s estimated to be $\log L_{bol}$ =  46.5$\pm0.05$ erg/s. The kinetic power carried by ionized outflows is about 0.01\% of the AGN bolometric luminosity, indicating that the feedback is likely not able to regulate the evolution of their galaxy hosts effectively \citep{Scannapieco2004,Hopkins2010}.

Throughout the paper, all reported uncertainties represent statistical uncertainties only. The statistical uncertainties from the spectral data are minimal compared to the systematic uncertainties.  We note that if we take the assumption that \oiii\ luminosity is ten times \hb\ luminosity, and calculate the \oiii\ mass by \hb\ following \citet{Nesvadba2011}, the mass of outflowing gas obtained is about seven times that of our result. If we adopt the electron density from our measurement n$_{e}$ = 272 cm$^{-3}$ using \oii, the outflow mass would be increased by a factor of two.
Although, the \oii\ ratio provides the average narrow line region density. This electron density value may not be the bulk density of the \oiii\ outflow. Narrow line region and outflow densities may significantly differ (e.g. \citealt{Brusa2015,Peng2015}). 
Nevertheless, the bulk of the outflow might be in a different phase (e.g. \citealt{Zubovas2012,Brusa2018,Travascio2024}). These estimates provide lower limits on the actual outflow rates, as only a fraction of the mass of the outflows is in the warm ionized phase traced by \oiii\ \citep{Cicone2018}. The outflowing gas in other phases (e.g. more highly ionized, molecular and atomic gas) is not traced by \oiii. Outflows in high-redshift radio galaxies in the molecular phase are typically 5 - 50 times more massive than ionized gas \citep{Nesvadba2017}. We do not consider the effect of the inclination and dust attenuation, either. Therefore, in the case the projection of the outflow velocity along the line of sight is small, the kinetic energy of the outflows measured from ionized gas could be underestimated.

\subsection{Comparison to absorption line analyses}  \label{sec:comtouv}

%In 3C 191, an associated absorption line (AAL) system with rich absorption lines was first discovered by \cite{Burbidge1966}. \citet{Hamann2001} observed 3C 191 using HIRES on the Keck I telescope. In our companion paper \citep{Sharma}, we analyze a more recent VLT/X-shooter observation of the outflow and also revisit the analysis of \citet{Hamann2001}, using up-to-date absorption analysis techniques. \textcolor{blue}{Therefore, we focus on comparing the IFS results to \citet{Sharma}.}
In our companion paper \citep{Sharma}, we analyze a more recent VLT/X-shooter observation of the outflow and also revisit the analysis of \citet{Hamann2001}, using up-to-date absorption analysis techniques. We focus on comparing the IFS results to \citet{Sharma}.

%\subsubsection{Outflow velocity} 
\subsubsection{Outflow velocity and extension}

Absorption line analysis reveals multi-components that are blueshifted by $\sim$ -500$\pm40$ to $\sim$ -950$\pm40$ km s$^{-1}$ in this system \citep{Sharma}. 
The velocity of our blueshifted emission outflow component in the nucleus (``B'' region) is $\sim$ -20$\pm50$ to -1090$\pm50$ km s$^{-1}$, which is consistent with the result of the absorption line, implying that they might have the same origin. These UV absorption lines are most likely generated by the part of the blueshifted gas falling in the line of sight.

Absorption line analysis yields an indirect method in measuring the distance of the outflow from the central source ($R$).  This is done through the use of absorption troughs from ionic excited states. The column density ratio between the excited and resonance states yields the outflow number density. Combined with knowledge of the ionization parameter of the outflow, $R$ can be determined. (e.g., \citealt{Arav2018}). For 3C 191, \citet{Sharma} use the \siiii\ and \siii\ troughs. From the X-shooter data, this analysis yields $R = 5.1^{+0.9}_{-1.0}$ kpc, in remarkable agreement with the size of the blueshifted outflow deduced from the IFS analysis (see more in Sharma et al. in preparation).  Note that there is a large systematic uncertainty in the distance estimated by absorption lines. For example, the original estimation using Keck data by \citet{Hamann2001} of $R = 28$ kpc differs a lot from our re-analyzing result of $R = 9.3^{+1.8}_{-2.5}$ kpc using Hamann’s data and more modern techniques.

We directly measure the size of the \oiii\ bubble through the maximum size of \oiii\ corresponding to the SNR threshold. The result are shown in table~\ref{tab:scale}. The scale of the \oiii\ outflow at different SNR threshold is comparable.
Therefore, the analysis of UV absorption using high-resolution data yields spatial scales comparable to this study.

The outflow size obtained from this IFS study is observed in the projected plane, while the absorption lines are observed along the line of sight. The blueshifted outflow extends to a projected distance of 5 kpc obtianed from this study and reaches the same distance along the line of sight, as indicated by absorption analysis. These could suggest a biconical outflow with a high inclination relative to the line of sight and a large opening angle.

\begin{table}[htb]\footnotesize
\caption{Outflow scale measured from IFS at different SNR.}
%\hspace{-0.7in}
\begin{tabular}{ccc}
\hline
\hline
 SNR & scale (blueshifted) & scale (redshifted)   \\
\hline
 & kpc & kpc  \\
\hline
$\geq$ 1.5 &   6.3    & 11.5    \\  
$\geq$ 3 & 5.2 &   11.0   \\ 
$\geq$ 5  &    4.8 &  8.9    \\   
\hline
\hline
\end{tabular}
%\hspace{0.7in}
\label{tab:scale}
\tablecomments{The scale is a maximum size from the nucleus corresponding to given a SNR threshold.}
\end{table}

\subsubsection{Outflow kinematics}

\begin{table*}[htb]\footnotesize
\caption{Gas kinematics measured from absorption line and IFS.}
%\hspace{-0.7in}
\begin{tabular}{cc|ccccc}
%\begin{tabular}{|c|c|c|}
\hline
\hline
   & & Abs & IFS (blueshifted) & IFS (redshifted) & IFS (bubbles) & \\
\hline
Parameter  & & Value & Value & Value & Value &Unit\\
\hline
velocity range & (1) &   -950 to -500    & -1090 to -20  &  20 to 820 & - & km s$^{-1}$\\  
$V_{med}$ & (2) & -720 &   -657$\pm41$ & 521$\pm30$ & - &  km s$^{-1}$\\  $V_{max}$ & (3) &  &   -975$\pm46$& 807$\pm50$ & -   &  km s$^{-1}$\\ 
R         & (4)   &    5.1$^{+0.9}_{-1.0}$ &     5$^{+1.3}_{-0.2}$  & 11$^{+0.5}_{-2}$  &  - &  kpc \\   
$t_{dyn}$    & (5) &    6.9$^{+1.2}_{-1.3} \times 10^{6}$  & 5.0$\pm0.3 \times 10^{6}$    & 1.3$\pm0.1 \times 10^{7}$ & - &  yr \\ 
$M_{\rm gas}$    &  (6)    & 2.3$\pm1.1 \times 10^{8}$  & 3.5$\pm0.6 \times 10^{7}$ & 3.1$\pm0.3 \times 10^{7}$  & 6.6$\pm0.7 \times 10^{7}$  &  $M_{\odot}$ \\
$\dot{M}$     & (7)          & 33$^{+13}_{-12}$ & 7.1-9.4 & 2.4-4.0 & 9.5-13.4 & M$_\odot$ yr$^{-1}$ \\
$E_{\rm kin}$   & (8)      & 1.2$\pm0.6 \times 10^{57}$   & 3.3$\pm0.8 \times 10^{56}$ & 2.0$\pm0.2 \times 10^{56}$ & 5.3$\pm0.9 \times 10^{56}$   &  erg \\
$\dot{E}_{\rm kin}$  & (9)      & 5.5$^{+2.1}_{-2.0}\times 10^{42}$  &(2.1-2.8) $\times 10^{42}$  & (4.8-8.6) $\times 10^{41}$ & (2.6-3.7) $\times 10^{42}$ & erg s$^{-1}$ \\
\hline
\hline
%  &   & IFS-\hb\ (blueshifted) & IFS (redshifted) & IFS (bubbles) & \\
%\hline  
%$M_{\rm gas}$          &  2.5$\pm0.8 \times 10^{8}$  & 9.2$\pm1.4 \times 10^{7}$ & 8.5$\pm0.7 \times 10^{7}$  & 17.6$\pm1.4 \times 10^{7}$  &  $M_{\odot}$ \\
%$\dot{M}$                & 34$^{+8}_{-7}$ & 8.5-12.0 & 5.6-6.3 & 13.4-19.0 & M$_\odot$ yr$^{-1}$ \\
%$E_{\rm kin}$          & 13$\pm4 \times 10^{56}$   & 3.2$\pm0.7 \times 10^{56}$ & 2.8$\pm0.6 \times 10^{56}$ & 6.1$\pm1.5 \times 10^{56}$   &  erg \\
%$\dot{E}_{\rm kin}$         & 5.6$^{+1.3}_{-1.1}\times 10^{42}$  &(0.8-1.2) $\times 10^{42}$  & (0.6-0.7) $\times 10^{42}$ & (1.5-2.1) $\times 10^{42}$ & erg s$^{-1}$ \\
%\hline
\end{tabular}
%\hspace{0.7in}
\label{tab:tab5}
\tablecomments{Gas kinematics measured by \citet{Sharma} using absorption line observed by VLT/X-shooter and our IFS data in this work, respectively. The velocity of the absorption line is calculated according to the redshift in this work. (1)Velocity range of the line profile. (2) The average value of $V_{med}$ from Figure ~\ref{fig:fk}. (3) $V_{max}=|\rm v_{med}| + 2\sigma$, which can be considered as representative of the bulk velocity of the outflowing gas. We measure the average value of $V_{max}$ of the blueshifted and redshifted from their velocity and velocity dispersion map (Figure ~\ref{fig:fk}). (4) We directly measure the size of the \oiii\ outflow through the maximum size of \oiii\ corresponding to the SNR threshold. (5) Dynamical timescale of the outflow. (6) Outflow mass. (7) Mass outflow rate. (8) The kinetic energy of the outflowing gas. (9) Kinetic luminosity of the outflow.}
\end{table*}

 We have shown that the outflow traced by the absorption lines is likely to be part of the blueshifted superbubble located in the line of sight. In a companion paper (Sharma et al, in preparation), we analyze the absorption data of the outflows based on observations with VLT/X-shooter. The kinematics measured by absorption line and IFS emission line are tabulated in table~\ref{tab:tab5}. We find a good agreement between the physical parameters of the outflow (distance from the central source $R$, $\dot{M}$ and $\dot{E}_{\rm kin}$), deduced from the IFS and absorption analyses.
The rough consistency between emission-line and absorption-line studies confirms both are effective approaches for characterizing outflow properties.

%The advantage of the absorption line analyses is that the density is measured accurately. However, the disadvantage is that the outflow radius is measured indirectly, and the global covering factor relies on assumption. The advantage of the IFS emission-line study is that the outflow radius is directly measured, and no covering factor needs to be assumed. 
Our results combine the advantages of these two outflow indicators and are, therefore, more reliable.
\subsubsection{Emission and absorption line in other sources} 

So far, limited works have been done to compare the kinematics of the outflow analyzed by absorption and emission line. \citet{Liu2015} obtained the IFS maps of \oiii\ outflow in two Seyfert galaxies at low-redshift, and thus presented one of the first such comparisons between AGN outflows analysed using absorption and IFS techniques. The IFS analysis of both these objects of the outflows were consistent with those from the absorption line analyses, including the scale, velocity, kinematics of the outflow. However, the velocity and the spatial extent of the \oiii\ outflow around HE 0238-1904 detected with IFS observation \citep{Zhao2023} is different with absorption line analyses \citep{Muzahid2012,Arav2013}. This strongly indicates that the outflows detected using the absorption and emission lines are clearly not the same component but likely stratified components of different spatial scale and velocity in the ionized phase outflow.

\subsection{Comparison with other outflows}

\begin{figure*}[htb]
\centering
\includegraphics[width=0.45\textwidth]{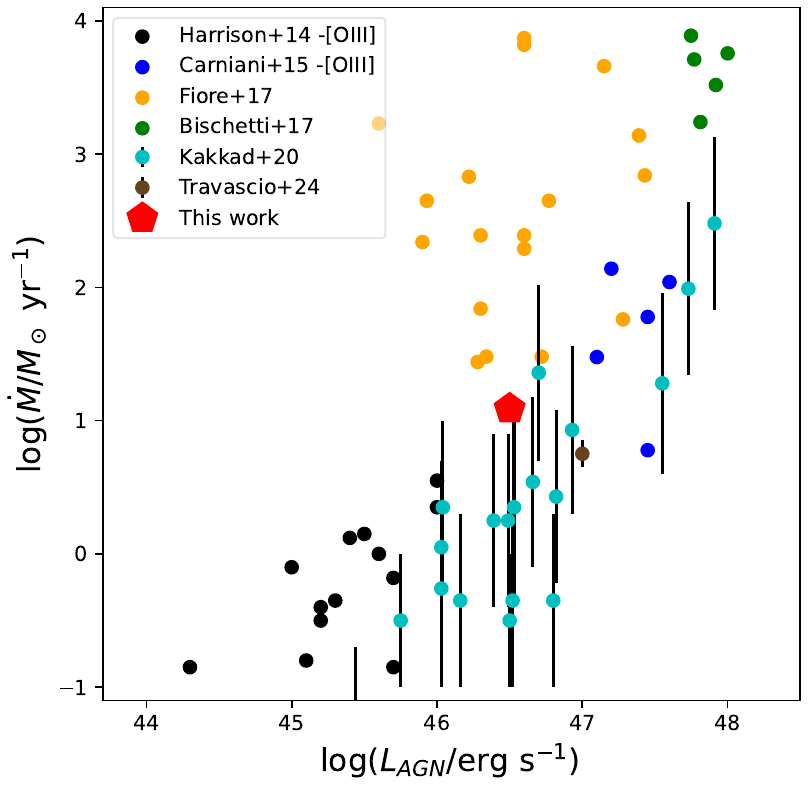}
\includegraphics[width=0.45\textwidth]{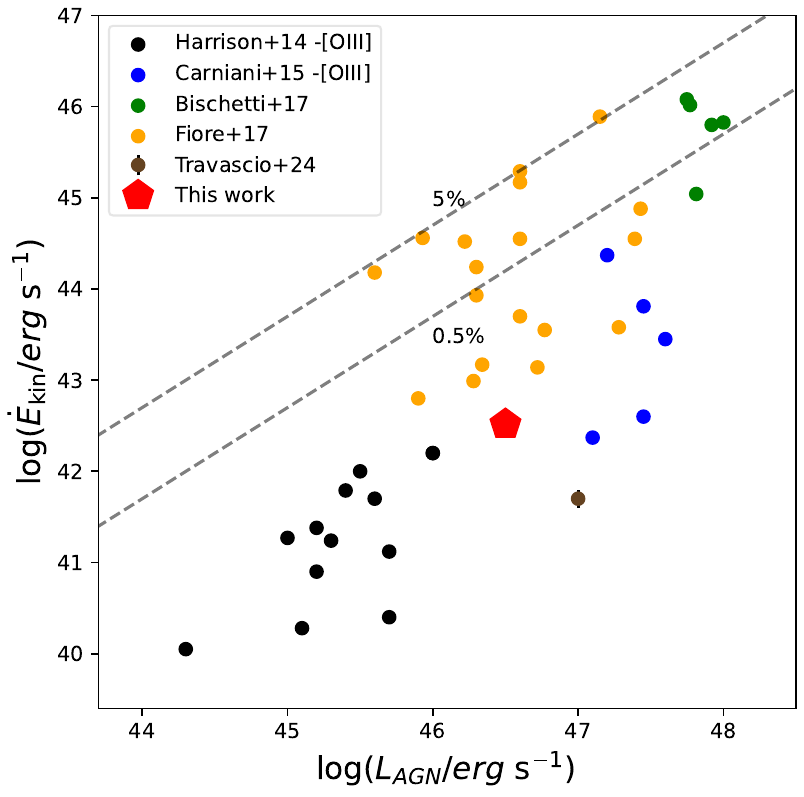}
\caption{Left: outflow rates as a function of the AGN bolometric luminosity. The red pentagon denotes the result from this work. The circles mark the outflow rates of ionised gas from different studies: black for \citet{Harrison2014}, cyan for \citet{Kakkad2020}, orange for \citet{Fiore2017}, blue for \citet{Carniani2015}, green for \citet{Bischetti2017}, and brown for \citet{Travascio2024}. Right: kinetic power as a function of the AGN bolometric luminosity. Symbols and colours are the same as shown in the left panel. The dashed lines correspond to P$_{k}$ = 5\%, 0.5\% L$_{AGN}$, respectively.
\label{fig:com}}
\end{figure*}

In Figure ~\ref{fig:com} we plot the outflow mass rate and kinetic luminosity as a function of the AGN luminosity. 
The black points represent the ionized outflows observed in type 2 AGN at redshift $0.08 \lesssim z \lesssim 0.2$ \citep{Harrison2014}. The brown point represents the \oiii\ outflows observed in the most luminous radio-quiet quasar at redshift $z \sim 0.2$ \citep{Travascio2024}.
We also include measurements from type 1 quasars and luminous AGN at $z \sim 2$ \citep{Kakkad2020,Fiore2017,Carniani2015,Bischetti2017}. In Figure ~\ref{fig:com}, we find that the outflow rate and kinetic luminosity are correlated with the AGN luminosity, although with a large scatter. Both the outflow rate and kinetic luminosity increase with increasing AGN luminosity. These are consistent with the theory that AGN inflate gas through the AGN radiation. Compared with pervious work, the kinetic luminosity of 3C 191 is smaller than those of other AGN at fixed AGN luminosity. This discrepancy could be due to the different n$_{e}$ values used in their studies. In this work, we adopt the electron density $n_e$ = 550 $\mathrm{cm}^{-3}$, while the \citet{Bischetti2017} and \citet{Fiore2017} assumed the electron density $n_e$ = 200 $\mathrm{cm}^{-3}$.  The outflow mass rate and kinetic luminosity are multiplied conservatively by a factor of 3  \citet{Bischetti2017,Fiore2017,Kakkad2020}. In addition, this discrepancy could be due to the type of AGN as the estimation of the outflow parameters is based on assuming a very simplistic scenario, and these parameters do not take into account the effects of inclination, opening angle of outflow, etc.

\subsection{Origin of the \oiii\ outflow}

AGN feedback is thought to operate in two main modes. In the jet mode, in which AGN feedback occurs through powerful ($\gtrsim 10^{45}$ erg s$^{-1}$) jets launched directly from the central AGN \citep{Fabian2012}. Simulations indicate that jets can drive outflows through interactions with the interstellar medium (ISM), generating shocks that spread outward (e.g. \citealt{Sutherland2007}). In this case, the jet-mode outflows are expected to extend along the direction of the radio jet (e.g. \citealt{Feruglio2020,Ulivi2024}). Different PA for radio jets and \oiii\ winds are often observed (e.g. \citealt{Husemann2019}), while other times they are almost aligned \citep{Jarvis2019,Shimizu2019,Feruglio2020}. To examine the optical \oiii\ emission and any relationship with the jet, we show the \oiii\ emission contours over the VLA radio contours and Chandra ACIS-S X-ray emission contours in Figure ~\ref{fig:png}. For 3C 191, the \oiii\ map is preferentially aligned along the radio jet direction and spans 2.25$\arcsec$, almost half that of Chandra ACIS-S images \citep{Sambruna2004}. The \oiii\ $\lambda$ 5007 is extended, showing an anisotropic distribution. The extended \oiii\ tends to be brighter on the side of the nucleus with the stronger, jet-like radio emission. 
However, observationally, the axis of \oiii\ outflows in this work is projected in a roughly northeast-southwest direction, with a PA of $\sim$ 330$^{\circ}$. There is a angular difference between the axis of \oiii\ outflow and the radio jet with PA of $\sim$ 10$^{\circ}$ (Figure ~\ref{fig:png}). Therefore, we rule out jets as the primary driver of the \oiii\ outflows.

The second mode is the quasar mode, in which the radiation emitted by AGN drives powerful outflows (e.g. \citealt{Liu2013a,Shen2023}). BAL outflows likely originate from the torus \citep{He2022}. Simulations show that the BAL outflows launched at the torus can propagate into the ISM and produce inflated bubbles \citep{Shen2023}. A recent study concludes that no significant difference is seen in the \oiii\ emission-line properties between the BALs, non-BALs, and mini-BALs at z$\approx$2 \citep{Temple2024}. This implies that outflow around 3C 191 is likely to launch at the torus and inflate bubbles. Part of the blueshifted gas fall in the line of sight, thus the blueshifted gas may manifest itself as absorption lines in the UV spectrum. The velocities and size of the outflow measured from the \oiii\ emission line and UV absorption lines are consistent, indicating a common origin. We conclude that the outflow found in 3C 191 is likely driven by radiation.

%\textcolor{red}{The blueshifted outflow projected extend to 5 kpc, while the gas moving towards us can produce absorption at 5 kpc along the line of sight. This may due to the bicone outflow with high inclination with respect to the line of sight and large opening angle. The actual velocity and the size of the blueshifted and redshifted outflow may be inconsistent.}
\section{Summary}

This paper presents IFS observations using VLT/SINFONI aided by AO in the near-IR band of a radio-loud quasar 3C 191 at redshift $z \sim$ 2.

The \oiii\ emission comprises three components separated in position and velocity: a blueshifted component in the north, a redshifted component in the south, and an unshifted component in the center with bounded motion in the quasar host galaxy.
The blueshifted and redshifted \oiii\ emission likely form a pair of outflowing superbubbles, mostly likely driven by AGN. 
The outflows of 3C 191 extend to at least 5 kpc and 11 kpc to north and south from the center, and have radial velocities of $\sim 975\pm46$ km s$^{-1}$, $\sim 807\pm50$ km s$^{-1}$, respectively.
The dynamical timescale of the gas can be estimated at $\sim 5.0\pm0.3 \times 10^{6}$, $\sim 1.3\pm0.1 \times 10^{7}$ yr as the travel time to reach the observed distances from the center.
The unshifted component that has bounded motion in the quasar host galaxy is consistent with the typical narrow line region in AGNs.

The outflow traced by absorption lines, as investigated in previous works, is likely to be part of the blueshifted superbubble because of the similarities in velocity and size.
We adopt the size measured from IFS data, and electron density $n_{e} = 602 $ cm$^{-3}$ measured from absorption lines, and estimate the mass of the outflow to be $M_{\rm gas} \sim 6.6\pm0.7 \times 10^{7}~ M_{\odot}$, and the mass outflow rate to be $\dot{M} \sim $ 9.5-13.4 M$_\odot$ yr$^{-1}$. Then, the kinetic energy carried by the ionized gas is 5.3$\pm0.9 \times 10^{56}$ erg, corresponding to a rate of $\dot{E}_{\rm kin} \sim$ 2.6-3.7 $\times 10^{42}$ erg s$^{-1}$. Only $\sim$ 0.01\% of the bolometric luminosity of 3C 191 converted to the kinetic power of the ionized outflow, which is far less than the value required for significant feedback derived from theoretical works ($\sim$0.5\% – 5\%) \citep{Scannapieco2004,Hopkins2010}.
These results suggest that the feedback of the ionized outflow may take place on kiloparsec or galaxy-wide scales, but is inadequate to regulate effectively the evolution of their galaxy hosts.

This work bridges emission- and absorption-line analyses of the ionized phase of AGN outflows and determines the outflow properties, in line with our previous work that conducts a similar combined analysis of two low-redshift quasars \citet{Liu2015}. Our results confirm the galactic-scale outflow traced by absorption lines and provide more reliable outflow radius and mass measurements.
Our analysis shows that the outflow parameters can be better constrained by combining emission and absorption probes.

%% Also note that the akcnowlodgment environment does not support long amounts of text. If you have a lot of people and institutions to acknowledge, do not use this command. Instead, create a new \section{Acknowledgments}.
\begin{acknowledgments}

Based on observations collected at the European Organisation for Astronomical Research in the Southern Hemisphere under ESO programme(s) 097.B-0570(B) and 092.B-0393(B). This paper employs a list of \emph{Chandra} datasets, obtained by \emph{Chandra} X-ray Observatory, contained in the \emph{Chandra} Data Collection (CDC) 289 \dataset[doi:10.25574/cdc.289] {https://doi.org/10.25574/cdc.289}. This research has made use of the software provided by the Chandra X-ray Center (CXC) in the application packages CIAO and Sherpa.

We acknowledge the research grants from the China Manned Space Project (the second-stage CSST science project: {\em Investigation of small-scale structures in galaxies and forecasting of observations}), the National Natural Science Foundation of China (No. 12273036), the Ministry of Science and Technology of China (National Key Program for Science and Technology Research and Development, No. 2023YFA1608100),  and the support from Cyrus Chun Ying Tang Foundations. Q.Z. acknowledges the support from the China Postdoctoral Science Foundation (2023M732955). NA acknowledges support from NSF grant AST 2106249, as well as NASA STScI grants AR-15786, AR-16600, AR-16601, and AR-17556.
\end{acknowledgments}

\begin{figure*}[htb]
\centering
\includegraphics[width=0.9\textwidth]{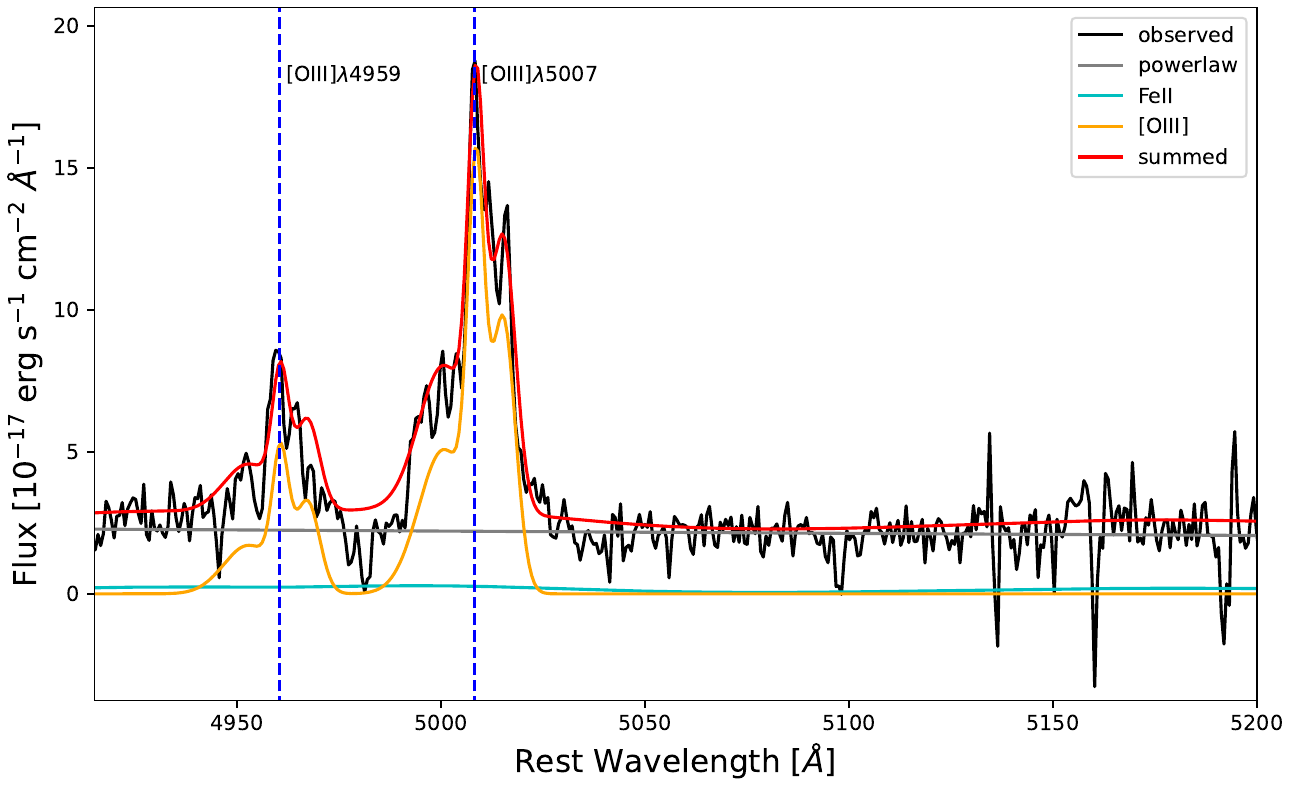}

\caption{The initial spectrum of the 3C 191 extracted from the field of view of $2'' \times 2''$ (40 $\times$ 40 pixel) observed by VLT/SINFONI. The observed spectrum (black) and the best-fitted spectrum (red) are shown. The different components of the best-fitted model are shown as different colored lines: the power-law component in grey,  the \feii\ component in cyan, and the \oiii\ component in orange. The zero velocity
are marked by blue dashed lines.
\label{fig:ins}}
\end{figure*}

%\begin{appendices}
\appendix
Figure ~\ref{fig:ins} shows the spectra extracted from the field of view of $2'' \times 2''$ (40 $\times$ 40 pixel) observed by VLT/SINFONI.

%\end{appendices}





\bibliography{./re.bib}{}
\bibliographystyle{aasjournal}



\end{document}